# Theoretical Analysis of the Structure, Thermodynamics, and Shear Elasticity of Deeply Metastable Hard Sphere Fluids

Subhashish Chaki[1,4,5], Baicheng Mei[1,4] and Kenneth S. Schweizer[1-4]

[1] Departments of Materials Science, [2] Chemistry, and [3] Chemical & Biomolecular Engineering,

[4] Materials Research Laboratory, University of Illinois, Urbana IL 61801 USA

[5] Institut für Theoretische Physik II–Soft Matter, Heinrich-Heine-Universität, Düsseldorf, Germany

Corresponding authors:
 kschweiz@illinois.edu, subhasischaki@gmail.com, bcmei@illinois.edu

**Abstract**

The structure, thermodynamics and slow activated dynamics of the equilibrated metastable regime of glass-forming fluids remains a poorly understood problem of high theoretical and experimental interest. We apply a highly accurate microscopic equilibrium liquid state integral equation theory that has not been previously explored in the deeply metastable regime, in conjunction with naïve mode coupling theory of particle localization, to study in a unified manner the structural correlations, thermodynamic properties, and dynamic elastic shear modulus in deeply metastable hard sphere fluids. New and distinctive behaviors are predicted including divergent inverse critical power laws for the contact value of the pair correlation function, pressure, and inverse dimensionless compressibility, and a splitting of the second peak and large suppression of interstitial configurations of the pair correlation function. The dynamic elastic modulus is predicted to exhibit two distinct exponential growth regimes with packing fraction that have strongly different slopes. These new thermodynamic, structural, and elastic modulus results are consistent with simulations and experiments. Perhaps most unexpectedly, connections between the amplitude of long wavelength density fluctuations, dimensionless compressibility, local structure, and the dynamic elastic shear modulus are discovered. These connections are more broadly relevant to understand the slow activated relaxation and mechanical response of colloidal suspensions in the ultra-dense metastable region and deeply supercooled thermal liquids in equilibrium.

## I. Introduction and Theoretical Background.

The hard sphere (HS) model is characterized solely by generic excluded volume repulsive interactions. For over half a century, it has been the foundational model for understanding structure and dynamics in dense correlated liquids, concentrated colloidal suspensions, and granular matter [1–6]. Moreover, it serves as the reference system for predicting the thermodynamic consequences of attractive interactions, such as within the perturbation framework of the modern van der Waals approach [3]. Monodisperse hard spheres (diameter, $\sigma$) are fluids up to a packing fraction of $\phi \equiv \frac{\pi \rho \sigma^3}{6} \sim 0.495$ ($\rho$ is particle number density), beyond which they crystallize in equilibrium. However, when crystallization is avoided for any reason, the HS fluid becomes a foundational model of metastable glass-forming matter, a topic of broad interest in condensed matter physics, physical chemistry, materials science, and biophysics. Disordered monodisperse smooth hard spheres achieve maximum random close packing (RCP) at $\phi_{RCP} \approx 0.64$ where mechanical jamming occurs and the pressure, contact value of the pair correlation function, and inverse compressibility all simultaneously diverge in a critical power law manner [4,5].

The structure of HS fluids is theoretically well understood in the "normal fluid" regime ($\phi \leq 0.495$) based on the Ornstein-Zernike (OZ) integral equation [7]:

$$h(r) = C(r) + \rho \int d\vec{r'}\, C(|\vec{r} - \vec{r'}|) h(r') \qquad (1)$$

Here, $h(r) \equiv g(r) - 1$ is the non-random part of the pair correlation function $g(r)$, and the negative of the direct correlation function, $C(r)$, in thermal energy $k_B T$ units is akin to a density-dependent effective pair potential with many body contributions. The static structure factor in Fourier-transform space is given by $S(k) = 1 + \rho h(k)$. The equation of state (EOS) can be obtained from two formally exact statistical mechanical routes (virial and compressibility theorems) to the pressure which probe very different aspects of structure [7]:

$$\frac{\beta P}{\rho} \equiv Z = 1 + 4\phi g(\sigma) = \phi^{-1} \int_0^\phi d\phi' \, S_0^{-1}(\phi') \qquad (2)$$

Here, $g(\sigma)$ is the contact value of $g(r)$, and $S_0$ the dimensionless compressibility which obeys [7]: $S_0 \equiv S(k=0) = (1 - \rho C(k=0))^{-1} = \left[\frac{d\beta P}{d\rho}\right]^{-1} = \rho k_B T \kappa_T = V <(\delta\rho)^2> \rho^{-1}$, with $\beta \equiv (k_B T)^{-1}$, $\kappa_T$ the isothermal compressibility, and V the system volume.

Solving Eq.(1) requires an approximate closure relation. In the deeply metastable regime, new many body effects in C(r) become critical, and it has been argued that all OZ-based theories fail [5]. A widely employed empirical EOS, the Carnahan-Starling (CS) formula [8] $Z \equiv \beta P \rho^{-1} \approx (1 + \phi + \phi^2 - \phi^3)(1-\phi)^{-3}$, is quite successful, but breaks down beyond [4,5,9–11] $\phi \approx 0.56 - 0.57$. This has motivated much recent theoretical and simulation work [9,12–14] that claims there is a crossover in the *equilibrated deeply metastable* regime to a new equilibrium state, the mechanism for which remains debated [5].

Simulations find non-normal fluid behaviors emerge *quite far* from RCP including [4,5]: (i) critical power law growth of the contact value and pressure, $\beta P \rho^{-1} \propto g(\sigma) \propto (\phi_{RCP} - \phi)^{-1}$, (ii) splitting of the second peak of g(r), and (iii) power law vanishing of $S_0 \propto (\phi_{RCP} - \phi)^2$. It has been suggested these behaviors (and others) indicate a qualitative change of the free energy landscape, with perhaps an attendant configurational entropy driven phase transition [5]. The above (sometimes called "free volume") critical power law form [5,15] of the EOS can be deduced from a phenomenological cell model where particles are *assumed* to be localized in Voronoi cells in a globally disordered state [15], although objections to this perspective have been advanced [5].

The classic approximate Percus-Yevick (PY) and Hypernetted Chain (HNC) closures, though diagrammatically derived, do not properly capture points (i)-(iii) above and other features [5,16] in the deeply metastable regime. Recently, the hybrid modified-Verlet (MV) closure [17–19], defined as $g(r) = \exp[-\gamma(r) + b(r)]$ where $\gamma(r) \equiv h(r) - C(r)$ with the bridge function [17–19] $b(r) \approx -0.5\,\gamma^2(r)[1 + 0.6|\gamma(r)|]^{-1}$ for $r \geq \sigma$, has been studied up to $\phi \sim 0.6$. It combines HNC and PY ideas in a specific, albeit heuristic, manner and captures exactly the first 4 virial coefficients of the EOS. The OZ-MV predictions for g(r), S(k), structural metrics, and thermodynamic properties have all been recently shown to be in excellent agreement with crystal-avoiding simulations of monodisperse HS fluids up to $\phi = 0.585$ [20,21]. This includes unanticipated exponential growth laws with density [19] of the pressure, $S_0^{-1}$, collective order parameters, and the medium range order real space correlation length in the modestly metastable regime.

Other hybrid closures that mix the PY and HNC approximations in different heuristic ways also exist and have been shown to be accurate in the normal fluid regime. Perhaps most well-known is the Rogers-Young (RY) approximation which is technically different than the MV closure. For example, it invokes an adjustable parameter(s) in a model functional form for the direct correlation function outside the hard core which is chosen to enforce by hand thermodynamic self-consistency of a specific property [22]. This closure has not been applied, to our knowledge, in the (deeply) metastable regime. Our goal in the present article is not to study alternative closures. Rather, we employ the recently well-validated OZ-MV approach to explore in a unified manner the physics that determines the thermodynamics, structure, and shear elastic modulus in the deeply metastable regime at packing fractions beyond all prior studies.

The derivative of $k_B T C(r)$ plays a central role as an effective force in microscopic dynamical theories [7,23]. As both a further application and test of the OZ-MV theory of structure, and to address the puzzling experimental observation [24] that the elastic shear modulus of hard sphere colloidal suspensions exhibits two *exponential* growth laws with packing fraction separated by a crossover at $\phi \approx 0.6$, we combine the OZ-MV theory and a self-consistent mode coupling approach to analyze particle localization and dynamic elasticity in the metastable regime.

Our new thermodynamic and structural results are presented in section II, including the construction of a theoretical understanding of the aforementioned points (i)-(iii) above and the prediction of *two* distinct types of metastable structural and thermodynamic equilibrium behaviors. Section III shows that as a consequence of the novel structural predictions, a two regime exponential growth of the elastic shear modulus with packing fraction is predicted which is in agreement with experiment. Section IV present surprising new predictions for the inter-connections between thermodynamics, structure, and dynamic elasticity. The article concludes in Section V with a summary and future outlook.

**II. Structural and Thermodynamic Results.**

We first consider how g(r) evolves with density in the deeply metastable regime. Fig.1(a) shows that at intermediate packing fractions, such as $\phi = 0.55$, the basic features are akin to those of the normal fluid: a sharp decrease after the contact peak, followed by a minimum and a much smaller and "structureless" second peak. However, as packing fraction becomes sufficiently high (e.g., $\phi = 0.63$ as shown), g(r) exhibits a step-like near discontinuity at $r = 2\sigma$ and a change of shape of the second peak. At even higher $\phi$, the step-like feature becomes much sharper, a shoulder at $r \approx 1.6\sigma$ emerges corresponding to a split $2^{nd}$ peak, and the amplitude of the $3^{rd}$ peak strongly increases. Moreover, $g(r) \rightarrow 0$ in the window of interstitial-like interparticle separations of $r \sim (1.0 - 1.5)\sigma$. Fig.1b shows the contact value changes from well described by the phenomenological CS formula [$g(\sigma) = (Z-1)/4\phi \approx 0.25(1 + \phi - \phi^2)(1-\phi)^{-3}$] to $g(\sigma) \propto (\phi_{RCP} - \phi)^{-1}$ at $\phi \approx 0.62$.

The above features suggest that OZ-MV theory predicts a new type of amorphous order at very high $\phi$ of a more "solid-like" nature. Although some of these results may seem reminiscent of g(r) of a jammed packing (diverging contact value, split 2nd peak) [4,5], there are qualitative differences including the massive suppression of interstitial configurations, detailed form of the 2nd coordination shell peak, and large 3rd shell peak. These differences are expected since we are analyzing equilibrated fluids, and not literal mechanically jammed states.

Concerning the functional form of the contact value, Fig.1b shows that $g^{-1}(\sigma)$ exhibits a bi-linear decay in the metastable regime to a very high degree of numerical accuracy. The high-density branch is thus inferred to correctly obey the classic inverse critical power law scaling $g(\sigma) \propto (\phi_{RCP} - \phi)^{-1}$. The crossover between the two regimes occurs a bit higher than typically reported in simulations (which vary) [4,5,9,12–14,25], but still well below the packing fraction required for the clear emergence of a split 2nd peak in Fig.1a. This result supports our claim that OZ-MV theory predicts a new deeply metastable regime at $\phi > 0.6$. A small linear extrapolation to zero of our numerical calculations in Fig.1b yields $\phi_{RCP} = 0.776$. This is an unphysically high value, as expected given the non-existence of any microscopic liquid state theory that captures the exact location of the RCP state [26]. On the other hand, the obtained value is far closer to 0.64 [4] than $\phi_{RCP} = 1$ predicted by OZ-PY theory or the phenomenological CS formula. Moreover, and in contrast to OZ-MV theory, the latter two approaches predict *qualitatively* incorrect critical power laws for the pressure and contact value in the deeply metastable regime, $P \propto g(\sigma) \propto (1 - \phi)^{-x}, x = 2$ or 3 , respectively. Below we argue the error of OZ-MV theory for the value of $\phi_{RCP}$ does not invalidate its new predictions in the deeply metastable regime.

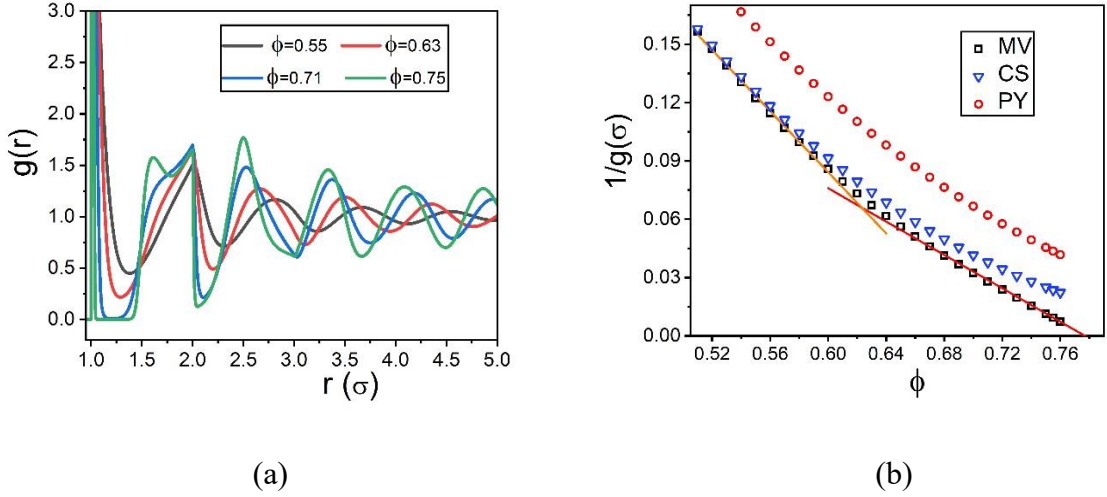

(a)                 (b)

Fig. 1. (a) Pair correlation function at 4 high packing fractions vs nondimensionalized interparticle separation. (b) Inverse contact value of g(r) versus packing fraction from OZ theory with the MV and PY closures, and the empirical CS formula. The two linear regimes predicted by OZ-MV theory, and the extrapolation to the RCP state, are indicated by the lines.

Figure 1b has immediate consequences for the EOS in Eq. (2). Fig. 2 shows the virial reduced pressure (first equality in Eq. (2)). Over a significant range of packing fractions in the deeply metastable regime, the pressure properly increases as a critical power law of a simple pole form. Simulations [4,5,19] find the CS formula begins to fail beyond $\phi \approx 0.56 - 0.57$, and Fig. S1 [27] demonstrates its deviation from OZ-MV theory emerges at a similar packing fraction. Though prior simulations [19] up to $\phi = 0.585$ found OZ-MV theory displays a high level of thermodynamic route consistency, its predictions for the pressure and its density derivative (which determines $S_0$) were demonstrated to be quantitatively most accurate when computed from the virial route [see Eq.(2)]. Moreover, they are much better than the corresponding results from the CS formula and the OZ-PY compressibility and virial routes.

The inset of Fig.2 plots the square root of the virial route determined dimensionless compressibility versus packing fraction. This format illustrates that OZ-MV theory properly predicts a quadratic critical power law, $S_0 \propto (\phi_{RCP} - \phi)^2$, from the crossover to the ultra-dense regime at $\phi \sim 0.62$ to the deduced RCP location which is the same as in Fig.1b. Also shown is the compressibility route analog, which is similar in all respects, except the deduced $\phi_{RCP}$ is slightly larger. These findings are significant since the PY and CS formulas predict a dimensionless compressibility that vanishes in a qualitatively much stronger manner with a qualitatively incorrect exponent, $S_0 \propto (\phi_{RCP} - \phi)^4$ where $\phi_{RCP} = 1$.

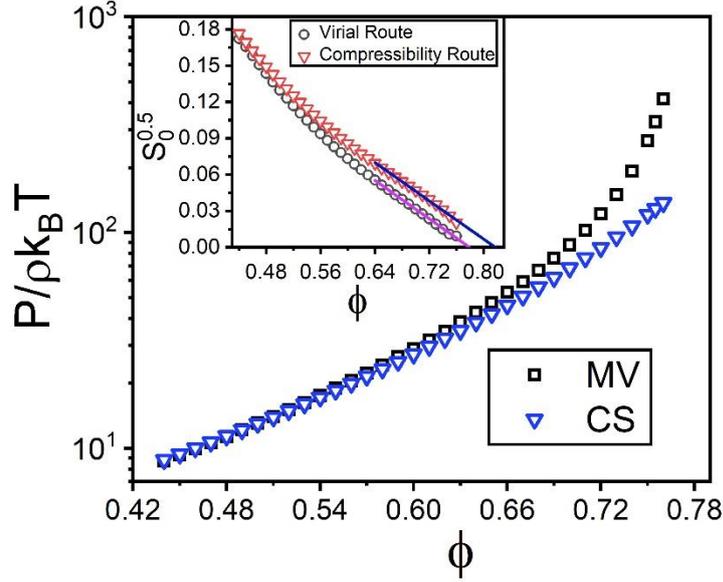

Fig. 2. Log-linear plot of the dimensionless pressure versus packing fraction from OZ-MV theory and the empirical CS formula. (Inset) Square root of the dimensionless compressibility as a function of packing fraction determined from the virial and compressibility routes of OZ-MY theory. Converged calculations were obtained down to very small values of $S_0 \cong 10^{-4}$.

The origin of the above OZ-MV theory results must be new features in the direct correlation function associated with emergent many body effects. Fig.3 shows representative results for $-C(r)$ in the first strongly metastable and the second "deep glass" regimes, defined as $\phi \leq 0.63$ and $\phi > 0.63$, respectively, which we will refer to as Regimes I and II. There are multiple key features. (i) Inside the hard core, $-C(r)$ in Regime II is very large, varies roughly linearly with particle separation, and tends to diverge as $r \to 0$ when $\phi \to \phi_{RCP}$. The latter feature is crucial for capturing the correct $S_0 \propto (\phi_{RCP} - \phi)^2$ behavior at high packing fractions since $S_0$ follows from the spatial integral of C(r) per the formula $S_0 =$ $(1 - \rho C(k = 0))^{-1}$. (ii) A sharp, delta-function-like peak emerges at a $\phi$−dependent position slightly beyond contact ($\frac{r}{\sigma} \sim 1.02 - 1.08$) in Regime II (see Fig.S2 [27]), which is the origin of the divergent critical power law form of $g(\sigma)$ and pressure in the deeply metastable regime. (iii) In both regimes, $-C(r)$ is negative beyond contact, indicating a many body attraction (inset of Fig.3). However, as previously shown [19], in Regime I this attraction is a maximum *at contact*, evolves modestly with density, and is very short range, only ~few percent of the particle diameter. In contrast, in Regime II the attractive tail amplitude has qualitatively new features: it attains a non-perturbative minimum at a *non-contact* interparticle separation of r~1.2σ when $\phi \geq 0.70$, and is a strong function of density. The overall range is far larger than in Regime I, though still short in an absolute sense since $C(r)$ approaches zero beyond r~(1.4 − 1.5)σ.

The above new features of $C(r)$ are critical for the emergence of a split 2$^{nd}$ peak and suppression of interstitial-like particle separations of g(r) in Fig.1a. Some are reminiscent of the $C(r)$ deduced from simulations of jammed packings using the OZ equation [4,28], but there are qualitative differences. For example, in jammed packings there is an asymptotic inverse power law scaling of $C(r) \propto -\sigma^2/r^2$, which is deeply related to the hyperuniformity of the structure factor, i.e., S(k → 0) approaches zero as a linear function of wavevector [4,28,29]. This is not predicted by equilibrium OZ-MV theory, consistent with the other qualitative differences in g(r) beyond contact discussed above in the context of Fig.1a.

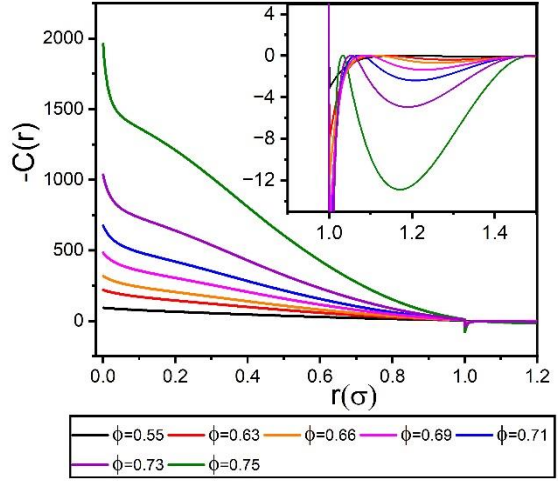

Fig. 3. Negative of the direct correlation function, inside (main) and outside (inset) the hard core, plotted as a function of interparticle distance (units of σ) for different packing fractions.

**III. Dynamic Elastic Shear Modulus**

The predicted new structural features in the metastable regime established above must play an important role in microscopic dynamics [7,19]. Here we study the dynamic elastic shear modulus, $G$, motivated by puzzling experiments for modestly polydisperse ultra-dense hard sphere colloid suspensions [24,30] with $\phi_{RCP} \approx 0.67$-$0.68$ [24,30]. Moreover, the emergence of particle localization and a dynamic shear modulus that persists to long times is of fundamental interest since it signals the transition from a fluid to a solid, or more generally a crossover to glassy dynamics.

The experimental colloidal suspension shear modulus data is shown in the inset of Fig. 4 for various solvents and colloid diameters [30], and superimposes well. Moreover, the experimentalists empirically discovered there are two regimes where $G$ exhibits a rather striking exponential increase with high slope parameters of $n \sim 42$ and 80, with a crossover at $\phi \approx 0.6$. There is no theoretical understanding of these observations.

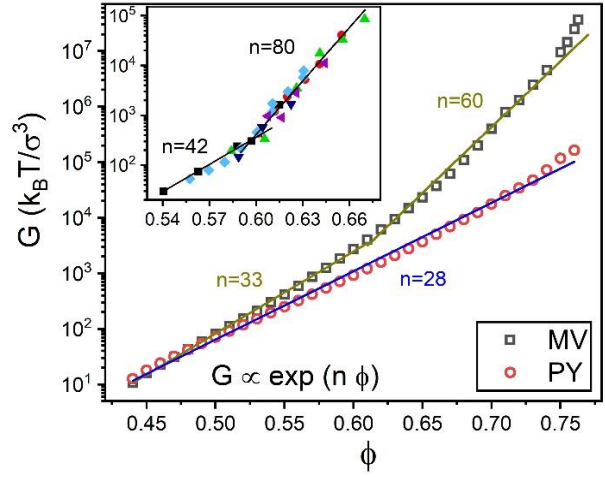

Fig. 4. Theoretical dimensionless elastic shear modulus versus packing fraction based on NMCT with the PY or MV closure employed for the required structural input. (Inset) Experimental dimensionless shear modulus as a function of packing fraction (see [30] for explanation of symbols).

We adopt the well-established naïve mode coupling theory (NMCT) [31,32] to predict the particle localization length ($r_{loc}$) and elastic shear modulus of an ideal glass relevant to the high frequency elastic modulus measurements [24]. The equations for the dynamic plateau shear modulus [24,33–36] and self-consistently determined localization length are [32]:

$$G = \frac{1}{60\pi^2} \int_0^\infty dk\, k^4 \left(\frac{d\ln(S(k))}{dk}\right)^2 \exp\left(-\frac{k^2 r_{loc}^2}{3S(k)}\right) \propto \phi k_B T \left(\sigma\, r_{loc}^2\right)^{-1} \quad (3a)$$

$$\frac{1}{r_{loc}^2} = \frac{1}{6\pi^2} \int_0^\infty dk\, k^4\, C^2(k)\rho S(k) \exp\left(-\frac{k^2 r_{loc}^2\left(1+S^{-1}(k)\right)}{6}\right) \quad (3b)$$

The key input is the Fourier space static correlations, and $r_{loc}$ which enters the single particle and collective Debye-Waller (DW) or nonergodicity factors modeled within an Einstein glass physical picture of the amorphous solid. Note that the collective DW contains the deGennes-like correction to a Vineyard-like form associated with static collective correlations. This simplified level of DW factor modeling is the only simplification of the full MCT approach adopted to compute the shear modulus in Eq.(3a). It has been successfully employed in diverse soft matter contexts for predicting the plateau dynamic shear modulus [32,34–36]. For the present deeply metastable hard sphere system, NMCT is expected to be especially reliable since the localization lengths are very small, and thus the wavevectors that dominate the arrested stress correlations in Eq. (3) are large. Hence, we expect none of our results will be significantly affected by including the more complicated collective contributions present in the full MCT formulation. The last proportionality in Eq. (3a) is an interesting consequence of the NMCT formulation. It relates the modulus to the localization length in a microrheological manner, as analytically derived and numerically shown in prior work [34], and explicitly verified here over the entire wide range of packing fractions studied (see Fig.S3 [27]).

The main frame of Fig. 4 presents our shear modulus results based on OZ-MV structural input, and the corresponding results using OZ-PY input. The predicted exponential increase of the elastic modulus with packing fraction is understood as a natural consequence of the previously obtained [37] NMCT result in Regime I that $\beta G \sigma^3 \propto \left(\frac{\sigma}{r_{\text{loc}}}\right)^2 \propto \left(\frac{\sigma}{\xi}\right)^6$, where $\xi$ is the structural correlation length which grows exponentially with packing fraction in both OZ-MV and OZ-PY theory [19]. However, use of OZ-PY structural input yields only a single exponential growth regime of G with a slope parameter of n~28. Up to $\phi$~0.60, the corresponding OZ-MV based results for the shear modulus are nearly the same, with a modestly larger value of n~33. Both calculations are in reasonable accord with the experiments [24] with regards to the first exponential growth regime up to $\phi \approx 0.60$. At higher packing fractions (Regime II), the elastic modulus determined using OZ-MV structural input exhibits a crossover at $\phi \approx 0.61$ to a much steeper exponential growth with a slope of n ~ 60. This crossover is missed based on using OZ-PY structural input, consistent with it is also not capturing all the key deep glass equilibrium features in the EOS, C(r), and g(r) analysed above.

Overall, we conclude that the OZ-MV plus NMCT results are in good agreement with experiment for (i) the existence of two exponential growth regimes, (ii) a slope ratio of ~2, and (iii) the location of the crossover, thereby providing the first explanation of these puzzling features. This agreement further supports our proposed 2nd crossover to an equilibrated "deep glass" Regime II at $\phi > 0.60$ since the second exponential growth law of G is a consequence of the corresponding equilibrium crossover predicted for the direct correlation function and g(r) in Figs. 1-3. This deduction is buttressed by noting that the predicted location and magnitude of the slope change of G is essentially identical to that predicted for the crossover of the pressure, $S_0$, and contact value to a critical power law form in Fig. 1(b), corresponding to a deep connection between elasticity, structure, and thermodynamics that we elaborate on in the next section.

Finally, Figure 4 hints that a 3rd narrow regime emerges very close to RCP where the theoretical elastic modulus exhibits modest upwards deviations from an exponential law. This is not unexpected since the hard sphere potential is singular and as RCP is approached $g(\sigma)$ diverges, and thus singular behavior of G can occur.

## IV. Thermodynamics-Structure-Elasticity Connections.

OZ-MV theory causally connects structure and thermodynamics, while NMCT is based on the idea of causal connection between elasticity, dynamics, and pair structure. Thus, one might expect our results for the shear modulus can be understood in terms of a dynamically appropriate metric of structural correlations and thermodynamics. This was indeed analytically and numerically verified previously based on the PY closure which found $r_{loc}^{-2} \propto \phi^2 g^4(\sigma)$, and hence $G \propto \phi r_{loc}^{-2} \propto \phi^3 g^4(\sigma)$, causal connections with a clear theoretical basis [38].

Figure 5 shows that a power law connection between the dynamic shear modulus and contact value based on the OZ-MV theory of structure is indeed predicted, and applies up to very high packing fractions (spanning 4 decades growth of G) with an apparent exponent of n=4.5. Far enough into the deep glass regime for $\phi \geq 0.71$, but still well below the theoretical RCP, there is a crossover to a different power law behavior with a much weaker apparent scaling exponent of n=2.4. This follows from the predicted qualitative changes of C(r) in the deep glass regime in Fig.3. Specifically, the emergence of a longer range attractive many-body tail with a *non-contact* minimum, along with the attendant split 2$^{nd}$ peak and suppression of interstitial-like interparticle separations in g(r) seen in Fig.1a.

A causal power law connection between the dynamic shear modulus and dimensionless compressibility, $G \propto S_0^{-3}$, has been recently established [37] in Regime I. Its origin is the predicted exponential growth with density of dynamically relevant equilibrium properties like $g(\sigma), \xi$, and $S_0^{-1}$ (see Fig.S4 [27]). This connection motivates the cross plot in the inset of Fig.5. One sees a power law scaling between the dimensionless shear modulus and inverse compressibility in Regime I holds over a wide range of high packing fractions. We emphasize this is *not* a literal dynamically causal connection since macroscopic thermodynamics does not determine localization and elasticity in NMCT. Rather it is a consequence of the predicted causal relationship between G and structure plus the equilibrium thermodynamics-structure connection [37,39].

A crossover in the very deeply metastable regime to a much weaker power law scaling between G and $S_0$ is also predicted. It can be understood based on the OZ-MV results that $P \propto g(\sigma) \propto (\phi_{RCP} - \phi)^{-1}$ and $S_0 \propto (\phi_{RCP} - \phi)^2$, which yields $G \propto g^{2.4}(\sigma) \propto S_0^{-1.2}$. The latter is consistent with the numerically established $G \propto S_0^{-1.3}$ scaling in the inset of Fig.5. We again emphasize that the most fundamental connection lies within NMCT which predicts $G \propto \phi r_{loc}^{-2}$ applies over all packing fractions in the wide metastable regime (see Fig.S3 [27])

Since the bulk modulus is given by $K_B = \rho k_B T S_0^{-1}$, the above $G - S_0$ relationship implies a power law proportionality between the elastic and bulk moduli, with different effective exponents in the two metastable regimes. In the very deep glass regime, the inset of Fig. 5 shows these two moduli are nearly proportional (n~1.3 close to unity). This is reminiscent of a continuum mechanics relation for isotropic solids [40]. Of course, the latter is not the system studied here, and the shear modulus studied here is a dynamically relaxed quantity, while the bulk modulus is not.

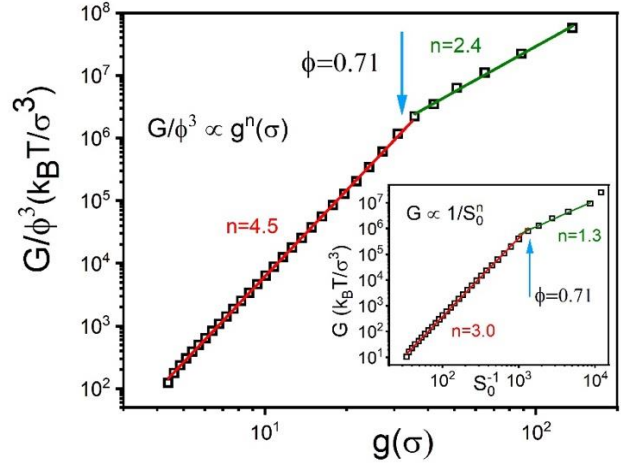

Fig. 5. Log-log plot of the dimensionless dynamic elastic shear modulus divided by the packing fraction cubed for all packing fractions studied plotted versus the (a) contact value (main frame), and (b) inverse dimensionless compressibility (inset).

## V. Summary and Broader Implications.

We have applied liquid state theory with a specific many body closure relation, previously well-validated [19] against simulations for monodisperse hard sphere fluids up to $\phi = 0.585$, to address the evolution of, and interrelations between, structural correlations, thermodynamics, and dynamic shear elasticity in the equilibrated deeply metastable regime. New predictions include: (i) inverse critical power laws as a function of $\phi_{RCP} - \phi$ for the contact value of g(r), pressure, and inverse dimensionless compressibility, (ii) distinctive features of g(r) including a splitting of second peak and anomalously low (high) interstitial (3rd coordination shell) spatial correlations, and (iii) two regimes of exponential growth with packing fraction of the dynamic shear modulus. All these results are consistent with simulations and/or experiments. Notably, the present findings emerge from an equilibrated microscopic statistical mechanical theory, in contrast with more phenomenological approaches and/or explanations that invoke, e.g., nonequilibrium jamming [4], solid-state-like cell models [15], or an unusual thermodynamic phase transition [5]. Unanticipated causal connections between the key *second* derivative fluctuation thermodynamic property that quantifies the amplitude of long wavelength density fluctuations ($S_0$), local structure ($g(\sigma)$), and dynamic shear elastic modulus, are also predicted. We believe our work provides evidence that the commonly expressed view that liquid state theory cannot capture unique features of the deeply metastable fluid regime is not true.

The emergence of hyperuniformity in $S(k)$ in mechanically jammed packing is not predicted for the equilibrated fluid model based on OZ-MV theory. An open future direction might be to investigate whether the long distance form of the closure, and hence C(r), can be adjusted to capture this feature without significant modification of all the other structural and thermodynamic behaviors properly captured by the existing MV closure.

A possible future application of OZ-MV theory is to employ it as structural input to either the non-self-consistent or the self-consistent versions of ideal MCT to analyze its consequences on the growth of the shear viscosity or inverse self-diffusion constant with packing fraction. However, in the deeply metastable equilibrated regime the importance of ergodicity-restoring activated processes seems unavoidably paramount [2,32,38], which are not accounted for in classic ideal MCT approaches.

Our findings, and OZ-MV theory more generally, are broadly relevant to activated dynamics and nonlinear rheology in deeply supercooled liquids and concentrated colloidal suspensions. For example, the present work sets the stage to address the nature of the structural relaxation time below the kinetic glass transition [41–45], or ultra-stable glasses deep on the landscape [46,47]. Key issues here include whether the alpha relaxation time diverges at a finite temperature or at a packing fraction below RCP, and whether there is a dynamic crossover *in equilibrium* to a relaxation mechanism of a simple Arrhenius form. These problems can be treated by combining successful microscopic force-based approaches for activated relaxation in glass-forming atomic fluids and colloidal suspensions [32,48] and polymer [37,39,49–51] and molecular [39,47,50] liquids using a predictive mapping under isobaric conditions to an effective hard sphere description [52], and the OZ-MV theory for structural correlations.

**Acknowledgement.** The authors acknowledge support from the Army Research Office via a MURI grant with Contract No. W911NF-21-0146. K.S.S. and B.M thank Professor George Petekidis for informative discussions of his experimental shear modulus data.